# Doppler-Shifted Alkali D Absorption as Indirect Evidence for Exomoons

**Carl A. Schmidt[1]**

[1]Center for Space Physics, Boston University, Boston, MA, USA

**\* Correspondence:**
Corresponding Author
schmidtc@bu.edu

**Abstract**

Sodium and potassium signatures in transiting exoplanets can be challenging to isolate from the stellar absorption lines. Here, these challenges are discussed in the framework of Solar System observations, and transits of Mercury in particular. Radiation pressure is important for alkali gas dynamics in close-orbiting exoplanets since the D lines are efficient at resonant scattering. When the star-planet velocity is $\gtrsim 10$km/s, eccentric exoplanets experience more than an order of magnitude higher radiation pressures, aiding atmospheric escape and producing a larger effective cross-section for absorbing starlight at the phase of transit. The Doppler shift also aids in isolating the planetary signature from the stellar photosphere's absorption. Only one transiting exoplanet, HD 80606b, is presently thought to have both this requisite Doppler shift and alkali absorption. Radiation pressure on a planetary exosphere naturally produces blue-shifted absorption, but at levels insufficient to account for the extreme Doppler shifts that have been inferred from potassium transit measurements of this system. In the absence of clear mechanisms to generate such a strong wind, it is described how this characteristic could arise from an exomoon-magnetosphere interaction, analogous to Io-Jupiter. At low contrasts presented here, follow-up transit spectra of HD 80606b cannot rule out a potassium jet or atmospheric species with a broad absorption structure. However, it is evident that line absorption within the imaging passbands fails to explain the narrow-band photometry that has been reported in-transit. New observations of energetic alkalis produced by the Io-Jupiter interaction are also presented, which illustrate that energetic sodium Doppler structure offers a more valuable marker for the presence of an exomoon than potassium.

## 1    Introduction

Despite being only trace species, sodium and potassium can dominate the optical absorption in exoplanetary atmospheres due to their uniquely high efficiency for resonance scattering. Alkali metals are unique in the way they interact with light. Photon scattering rates for their D line transitions are enormous compared to all other UV/optical/IR transitions in atomic species (e.g., resonant scattered solar Na D photon flux is ~350x stronger than Lyman α). This enables alkali absorption to be readily detected in transit spectroscopy. Na I and/or K I have been reported by this technique in nearly 30 exoplanets, with additional cases sure to follow[1]. The field is consequentially transitioning to science guided by comparison and characterization, not simply detection.

Interpretation of the alkali features in transit spectra requires disentangling the absorption features of the planetary atmosphere from that in the stellar photosphere. The brightest host-stars yield the best signal to noise transit spectra. Independent analysis of multiple datasets have proposed the alkali absorption in the prototypical hot-Jupiters HD 189733b and HD 209458b (e.g., Redfield *et al.*, 2008; Huitson *et al.*, 2012; Wyttenbach *et al.*, 2015; Khalafinejad *et al.*, 2017 and Charbonneau *et al.*, 2002; Snellen *et al.*, 2008; Langland-Shula *et al.*, 2009; Vidal-Madjar *et al.*, 2011, respectively). However, the atmospheric diagnostics among studies at these bodies have varied widely, and at times has even conflicted. At HD 189733b, Fisher and Heng (2019) pointed out that high-resolution spectroscopy of the Na doublet is incapable of accurately retrieving pressures at the level being probed by the lines—even at the order-of-magnitude level. Recently, the findings of Casasayas-Barris *et al.* (2021) disputed past Na detections at HD 209458b, and they attributed erroneous reports to center-to-limb variation (CLV) in the stellar photosphere and deformation of the stellar line profiles due to the Rossiter-McLaughlin effect. They asserted that there is no evidence of absorption due to the HD 209458b atmosphere when these corrections are implemented appropriately, even with the VLT/ESPRESSO instrument, which is optimized to investigate such features.

## 2    Solar system analogs: the case for eccentric exoplanets

It is reasonable to consider the Solar System a test bed for accuracy and success in disentangling alkali D line signatures: targets are much brighter, the planetary geometry is known precisely and in general, atmospheres have already been characterized independently, sometimes even by an *in situ* spacecraft. Solar system moons ingressing or egressing eclipse offer probes of the alkali tangent column through the parent planet's atmosphere. Yet, even under these ideal circumstances, separating the stellar and planetary alkali features has proven challenging. Vidal-Madjar *et al.* (2010) and Arnold *et al.* (2014) both used lunar eclipses to probe the terrestrial atmosphere and reported Na absorption (i.e., "Earth as an exoplanet"). Yan *et al.*, (2015) subsequently pointed out that such absorption owed to the intrinsic CLV of the Fraunhofer Na line profiles---not in fact an imprint of telluric Na. Since then, this experiment has not proven a viable means to measure Earth's atmospheric sodium. The Na layer in Earth's mesosphere is deposited by micrometeor ablation and has a tangent column of ~$10^{11}$ atoms/cm$^2$ depending on season and location. The corresponding emergent absorption equivalent widths are 14m Å for D2 and 7mÅ for D1 (expected equivalent width for a given column density and temperature are described by, e.g., Brown and Yung, 1976). But this relative signal is far weaker when reflected off the Moon because the Na layer is only 5 km in thickness and the differential refraction sounds a very extended altitude range. Moreover, interpretation is not straightforward because the ~$10^9$ cm$^{-2}$ Na tangent column in the lunar exosphere appears in emission, even in full eclipse (Mendillo and Baumgardner, 1995). The lunar exosphere's brightness is quite variable depending on the Moon's plasmasheet location (Wilson et al. 2006; Hapgood, 2007) and this contaminates the telluric absorption during eclipses since in this geometry it appears near null Doppler shift.

Sunlight reflected off Jupiter's moons as they cross the planet's penumbral shadow has similarly been used to infer the presence of an absorbing sodium layer in the jovian stratosphere (i.e., "Jupiter as an exoplanet"; Montañés-Rodríguez *et al.*, 2015). However, it is surprising that Na in Jupiter's atmosphere would be first detected by such an indirect means (D line absorption signature

---

reflected off Ganymede at ingress) and so this interpretation prompts skepticism. First, Fraunhofer lines from Fe I (5883.82Å) and Ni I (5892.88Å) appear in their Fig. 4, confirming sunlit artefacts in what should be a pure jovian absorption spectrum. Second, the "W-shape" line profiles that they reported may be a natural artefact of jovian scattered light contamination. Jupiter's size and rapid rotation broaden Fraunhofer features in its reflected continuum, and while Jupiter itself is not within the aperture, it is amply bright to contribute considerable stray light in its vicinity. Scattered light originates approximately uniformly from the jovian disk when pointing is far from Jupiter, but the telescope tracks towards Jupiter when observing a moon entering eclipse, and the reflected Fraunhofer lines grow narrower and deeper because the scattered sunlight is dominated by an increasingly localized region near Jupiter's limb. The jovian atmospheric transmission is obtained by dividing a satellite spectrum in penumbra by its spectrum in sunlight, but if some stray jovian light is present, this yields a narrow Fraunhofer line profile divided by a broadened line profile, which produces the "W" shape reported. Na deposition from above by Io or micrometeors seems an attractive explanation for the absorption. However, the region sounded by the Montañes-Rodriguez et al. (2015) observation reached 30 mbar depth, pressures $>10^4$ stronger than where micrometeors ablate in Earth's mesosphere. The diffusion time to this depth is of order years, and well exceeds the Na photo-ionization lifetime. In the upper stratosphere, the Na D lines would be pressure-broadened (Burrows and Volobuyev, 2003). These absorption equivalent widths correspond to a dense optically thick Na column, and in this saturated state the true column density is lesser depending on how strong the pressure-broadening is. If 30 mbar is considered as the maximum viable pressure broadening, then this sets a lower limit on the Na column density there: $7 \times 10^{11}$ cm$^{-2}$ (Oza *et al.*, 2019). Such a column would scatter at least 100 kR of D2 emission while Jupiter's limb continuum is ~1 MR/Å (e.g, Brown and Yung, 1976; R denotes Rayleigh units). Thus, even this lower limit on the jovian Na column should be readily detectable via high-resolution emission spectroscopy, but it has not been observed. Io's Na column density is comparable and indeed this is perhaps the only Solar System body where alkali absorption is unambiguous in the spectrum reflected off a background moon (Burger *et al.*, 2001).

A more straightforward analog for exoplanetary absorption lines is the alkali-rich transiting solar planet: Mercury. Sodium is a major constituent of Mercury's surface-bound exosphere, if not the primary species. On rare occasions, Mercury's exosphere can be measured in absorption as it transits the solar disk. Such observations can provide excellent spectral, spatial and temporal coverage of an otherwise elusive target---ample photons are available from a modern telescope pointed at the Sun. The exosphere is dynamic and not spherically symmetric and so transit observations are valuable because line absorption can effectively provide an image of the exosphere's column density at all points above the terminator (Schleicher *et al.*, 2004; Potter *et al.*, 2013).

Figure 1 shows May 9 2016 measurements of "Mercury as an exoplanet." The transit was observed by the Fast Imaging Solar Spectrograph (FISS; Chae *et al.*, 2015) on the 1.6m Goode Solar Telescope at Big Bear Solar Observatory (GSO; Goode and Cao, 2012). This transit's duration was 7.5 hours, but was already in progress at sunrise, allowing only 2.5 hours of observation. Stable tracking was achieved between 16:06:07 and 18:38:55 UT, when the high-order adaptive optics system lost its lock onto the planet's silhouette near third contact. The FISS slit scanned over the planet's disk in 130 steps every 16 seconds. The scan's step size perpendicular to the slit was chosen to match the 0.16"/pixel plate scale along the slit. This creates a 41" by 21" field of view per scan, over which the Na doublet was measured at a resolving power of R $= 1.4 \times 10^5$.

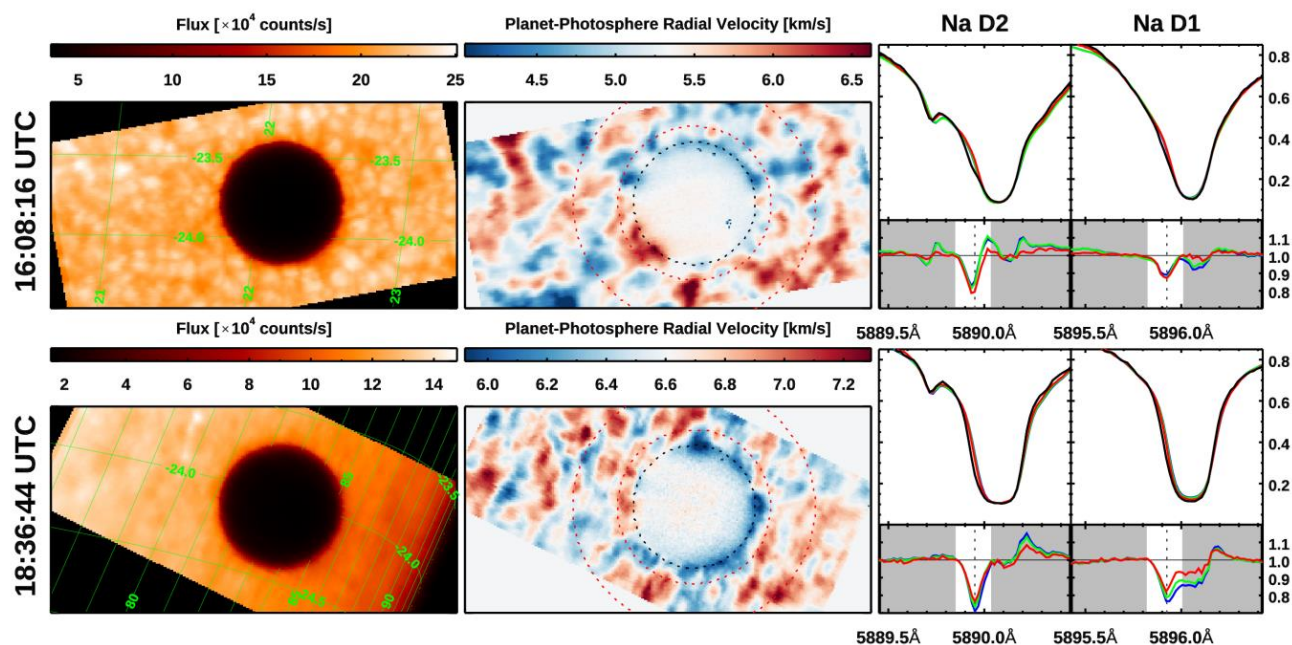

Figure 1. Intensity, radial velocity and Na D absorption line profiles during the 2016 transit of Mercury. **Left** panels show the wavelength-integrated count rates per spatial element, with Carrington heliographic coordinates overlaid. The solar limb is seen at bottom right, just prior to third contact. **Middle** panels show the radial velocity between Mercury's rest frame and measured line center of the Fraunhofer D lines. The innermost dashed ring outlines the planet's surface, and the two dashed red lines indicate the annulus applied in spectral analysis. **Right** panels shows an example spectrum (black) in a pixel at Mercury's equatorial dawn. Fitted reference spectra and their residuals are shown in color. Blue is the mean spectrum everywhere beyond the inner annulus ring. Green is the mean spectrum within the annulus bounds (red dotted lines) and co-aligned for Doppler shift. The red reference spectrum uses only the temporal information at this spatial location; it shows the interpolated spectrum at this solar coordinate from the nearest 6 scans in time where the solar photosphere was not occulted by the planet. Only grey spectral regions far from Mercury's line are used for fitting each reference spectrum, and white regions are used for integrating the exosphere's absorption.

The alkali distribution at Mercury's limb derived from this observation appears in Schmidt *et al.* (2018b) and Leblanc *et al.* (2022). The 27mÅ Na D2 equivalent width yields a tangent column of $1.6 \times 10^{11}$ atoms $cm^{-2}$ in the densest regions of the exosphere. Mercury's K tangent column is $\sim 10^9$ $cm^{-2}$, too thin to produce measureable absorption. However, it is the analysis itself that is relevant to exoplanetary science. In Figure 1, the absorption signal from Mercury's exosphere is isolated by constructing a reference solar spectrum to every spatial location in each scan using unobscured solar spectra obtained far from Mercury. Per the right panels, dividing the spectrum at a given location near Mercury's limb by a solar reference spectrum returns unity residual values, except where the Mercury exosphere absorbs. However, any reference spectrum only represents a best attempt to reconstruct how the solar spectrum at a given time and location would have appeared in the absence of Mercury. Analysis of Mercury transit data has explored different techniques for constructing optimal reference spectra. The most straightforward approach is to select all locations several atmospheric scale heights above the surface and use their mean spectrum as the reference (blue spectra, Figure 1). However, this approach averages over all Doppler structures present in the solar photosphere within the field of view (*cf.* middle panels). As in Potter *et al.* (2013), sampling an annulus around the planet and spectrally co-aligning them to the common Doppler shift of a given pixel produces residuals closer to unity (green spectra, Figure 1). Further improvements came by employing information in the neighboring datacubes of the time series (red spectra, Figure 1); a background solar longitude and latitude near Mercury's limb appears sufficiently far from the planet to be free of the exosphere's signal roughly a minute earlier and later, and with little change to the local spectrum over such a short duration.

At unresolved transiting exoplanets, Figure 1 illustrates several challenges that are also encountered when separating stellar and planetary alkalis features. At any given spatial pixel, the velocity structure in the solar Na layer (middle panels) limits how well planetary Na can be isolated. Thus, regardless of which of the above-described treatments of the local reference spectrum is applied, artefacts are always evident in the residuals at Fig. 1 right. These artefacts mostly average out over successive scans in the transit duration, but not completely. Exoplanets, of course, have no local reference spectrum. Only the stellar disk-integrated light is available empirically and so optimally recovering the local spectrum is non-trivial. Upper and lower panels in Fig. 1 span <1/3 of the solar disk at 24° S, yet the Fraunhofer Na line shapes shows strong variations over the solar disk: narrow near disk center and broad near the solar limb. Solar rotation imparts up to +2 km/s at each equatorial limb. This component, added with Mercury's heliocentric velocity, gives the true velocity between background and the absorbing gas (*cf.* middle panels). As the transit progresses, the total shift from these two components moves the relative wavelengths of solar and planetary absorption wells. Consequently, more sunlight is absorbed when Mercury and the local solar photosphere have a larger radial velocity. In Figure 1, Mercury's orbital eccentricity, *e* and its "argument of perihelion", *w*, gives the planet a ~4.5 km/s heliocentric velocity. The Sun's local radial velocity at the sight line location on the photosphere causes an additional Doppler shift of up to 2 km/s, effectively moving the planet's resonance absorption up and down one side of the solar Fraunhofer well. Thus, purely geometrical effects give Mercury's alkali absorption an asymmetric light curve during solar transits. This differs from the Rossiter-McLaughlin effect. The transit path in Fig. 1 is nearly co-planar with the solar rotation vector, but misalignment is common in exoplanets and the impact parameter is not always known (e.g., Bourrier *et al.*, 2020). In this case, line shape variations over the stellar disk, Doppler geometry and limb-darkening can all vary disproportionally, and if any element of the transit geometry is in question, the challenge of disentangling stellar and planetary alkalis becomes intractable.

Interpretation of transit spectroscopy from eccentric exoplanets is more straightforward and can circumnavigate many of the issues described. With sufficient planet-star velocities, the complexities produced by the local photosphere's Doppler shift and line shape variations over the stellar disk are irrelevant since the stellar continuum is absorbed as opposed to a stellar absorption well. A transiting planet occults a range of photosphere radial velocities that is within the stellar line profiles intrinsically, and so Doppler shifts where alkalis absorb the stellar continuum can mitigate complications that are otherwise imposed by the Rossiter-McLaughlin effect (e.g., Casasayas-Barris *et al.*, 2020). For Sun-like stars, the planet-star Doppler shift required to move from the line core into the continuum is >9 km/s for K and >20 km/s for Na, where the stellar lines are broader. Against the continuum, the flux absorbed by planetary alkalis is easier to isolate.

Radiation pressure is also stronger if the D lines are excited by the stellar continuum. The excitation frequency, and the radiation acceleration is given by Eqns. 6.2.4 and 7.2.9, respectively, in Chamberlain and Hunten (1989). At Mercury's semi-major axis of 0.37 AU, the anti-sunward acceleration on Na gas is 1.84 m/s$^2$ or half of Mercury's surface gravity (Schmidt *et al.*, 2010). At 0.05 AU from a G-type, continuum photons impart ~100 m/s$^2$ accelerations to Na atoms or four times Jupiter's surface gravity. Radiation pressure is about 50% greater at the K D-lines, and higher still in Li-D and the metastable Helium 10830Å triplet. Such forcing may affect dynamics in an exoplanet's upper atmosphere if these mixing ratios are high and the surrounding gas is light. Near the exobase, radiation pressure forces atoms into escape trajectories. Alkali lifetimes are vanishingly small near a parent star, but photo-ionization and radiation pressure both scale as inverse squares. Escaping Na would be ionized in 8 minutes on average (Huebner and Mukherjee, 2015) and be blue-shifted by 48 km/s when it disappears from the absorption spectrum. Photon momentum at re-emission also imparts recoil momentum to the atoms in a nearly random direction, thereby heating the cloud and growing its cross-sectional area relative to the stellar disk. In this way, eccentric exoplanets with appreciable star-planet radial velocities offer observers a significant advantage toward measuring and interpreting alkalis in exoplanetary atmospheres.

## 3 Exomoon signatures and Doppler shifted alkalis at HD 80606b.

HD 80606b is the only exoplanet where alkalis have been reported and the orbit is sufficiently eccentric ($e = 0.93$) that planetary alkali lines Doppler shift completely out of the stellar well to encounter the brighter continuum. Moreover, the planetary radial velocity is directed away from the star ($w = 301°$), so accelerating alkalis do not shift back into the stellar well. HD 80606b is a long-period (111.4 day), $3.94 \pm 0.11$ $M_J$ exoplanet that has been reported to have a 4.2% increase in the apparent planetary radius near the K D 1 line in transit (Colón *et al.* 2012). Radiation pressure and photo-ionization rates for K are both (coincidentally) 50% higher than Na (Potter, 2002). Extreme winds beyond the 121 km/s escape speed of HD 80606b were hypothesized as a possible explanation for observed color differences in the wings of the K line core (Colón *et al.,* 2012). Na velocities in Mercury's tail reach >100 km/s without being ionized (Schmidt *et al.*, 2012), and so it is worth considering if a rapidly escaping K exosphere could explain such high speeds.

Resonance scattering rates are high in the stellar continuum. At each scattering, the absorbed photon imparts its anti-stellar momentum and the re-emitted photon imparts its recoil momentum in a pseudo-random direction. In the exosphere, this effectively heats alkali gases and shapes their clouds into a tail. In primary transit, this comet-like tail is oriented towards the observer. Consequently, blue-shifted absorptions are observed in strong resonance lines (e.g., Wyttenbach *et al.*, 2015; Spake *et al.*, 2018; Borsa *et al.*, 2021). Again applying the formulas in Chamberlain and Hunten (1989) at the transit phase, the excitation frequency for the combined K D lines is 272 photons/atom/s corresponding to a radiation acceleration of $3.62$ m/s$^2$. This is well below the planet's $94.14$ m/s$^2$ surface gravity, and even if K had a mechanism for atmospheric escape, the gas would Doppler-shift only 12 km/s within its photo-ionization lifetime of just under an hour. Clearly radiation pressure is insufficient, so how might alkalis attain such high velocities? Again, our solar system offers an example: the moon-magnetospheric interaction at Io.

Building on the concept that Johnson & Huggins (2006) first put forward, Oza *et al.* (2019) and Gebek & Oza (2020) pointed out that alkali measurements at several exoplanets are consistent with extended clouds that could be produced by moon-magnetosphere interactions, like those of the Io-Jupiter system. At Io's orbit, the bulk flow velocity of plasma in the jovian magnetosphere is nearly co-rotational at 74 km/s, well above Jupiter's escape velocity. Ions in Jupiter's magnetosphere are neutralized via charge exchange or recombination, producing neutrals that are no longer frozen to the field and escape tangentially. These reactions at Io produce a diffuse Na nebula larger than the Sun (Mendillo *et al.*, 1990). While the distant cloud is symmetric, the fast alkali jet producing it is a highly directional feature local to Io's wake. With short photo-ionization lifetimes, the local jet is likely the sole surviving signature of any alkali neutralization at an exomoon. Jet velocity scales with the moon-planet distance and planetary rotation frequency. The HD 80606b rotation period is uncertain, but a model of its thermal response in Spitzer Space Telescope data suggests 93hr, much slower than predicted (de Wit *et al.,* 2016). If the HD 80606 magnetic axis were normal to the line of sight, a moon orbiting beyond $10^7$ km (152 planetary radii) could then eject the ≳200 km/s K velocities that Colón et al. proposed. This scenario seems unlikely: that exceeds the planet's periastron distance, an exomoon's orbital period would be comparable to the planet's period, and a slow-rotating planet may possess an insufficient dynamo for its magnetic pressure to stand off the stellar wind at such a distance. It is of course speculative to attribute highly red-shifted K absorption to ion neutralization in an exomoon-magnetosphere interaction. Yet, no other physical mechanism has been proposed to explain this observation. As moons and magnetospheres have both proven challenging to detect, HD 80606b is an interesting target for follow-up transit spectroscopy to characterize alkali features.

## 4    Transit spectra of HD 80606b

On 11-12 March 2019, in-transit and out-of-transit HD 80606b observing time was allocated at both the 4.3m Lowell Discovery Telescope and at the 3.5m WIYN telescope. Lowell observations were entirely weathered-out. WIYN targeted potassium only using the echelle grating of the Bench Spectrograph fiber fed from Hydra for a spectral resolving power R=15,000 at the K-D line doublet. This setup's 0.114 A/pixel dispersion equates to 4.4 km/s/pixel. The red fiber cable and STA1 CCD was used. Multiple fibers were aligned to stars in the nearby field for calibration and differential photometry. The full 12.1 hour transit duration of HD 80606b does not generally fit within a night and its ingress occurred in daylight during this measurement.

Only a few hours of WIYN data were obtained under cloudy conditions. The observing strategy was adapted when it became apparent that no out-of-transit spectra would be obtained due to weather: HD 80606b time in-transit was sacrificed to observe a reference star HD 80607. HD 80607 is a wide binary twin (angular separation 20.6″) that offers a very convenient control for this transit experiment, as it differs by only 0.1 mag. The binary also shares nearly identical chemical composition, with both being G5V spectral types and 80606 being only 0.013 dex more metal rich (Liu *et al.*, 2018). Moreover, it is through the differential photometry of these two stars that Colón *et al.* (2012) initially detected planetary potassium in transit. Ten 10 min spectra of each star where acquired, but HD 80707 in particular showed 2-3 magnitudes of cloud extinction. In the 7699Å D1 line, this limited dataset showed a Poisson S/N of ~245/pixel in HD 80606 and ~115/pixel in HD 80607. In Figure 2, red is HD 80607 black is HD 80606 near HD 80606b's mid-transit. Telluric $O_2$ poses formidable problems for the K D2 line, and surrounding artefacts remain after applying the transmission model of Leet et al. (2019) to account for differential airmass. The top panel is HD 80606/80607, in which the residual standard deviation around the potassium D1 line is ~1/90.

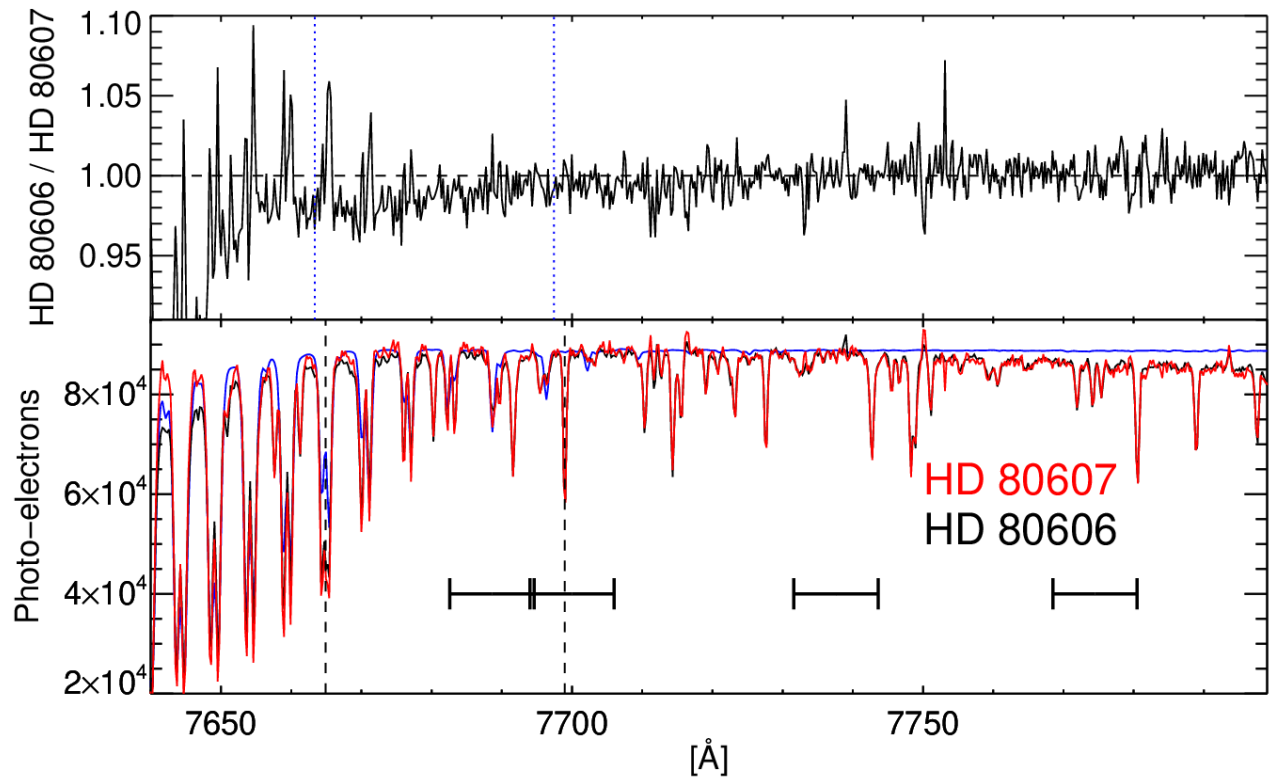

Figure 2. WIYN Hydra spectrum of HD 80606 during the center portion of an HD 80606b transit (black). A HD 80607 spectrum has been scaled to match dynamic range (red). Modelled telluric transmission is shown in blue, scaled to the continuum around K D1. The tunable filter bandwidths sampled by Colón *et al.* (2012) are shown as black bars. Black dashed lines indicate wavelengths of the stellar potassium D lines. The top panel shows the ratio of HD 80606 / HD 80607, where blue dotted delineates potassium lines blue-shifted by the eccentric planetary orbital motion.

Additionally, Keck/HIRES has made precision radial velocity measurements of the HD 80606 stellar wobble due to its eccentric companion. The Na D lines overlap with the iodine cell used in this setup, but potassium D1 is uncontaminated. Figure 3 shows archival data from the Winn *et al.* (2009) radial velocity measurements. Red shows out-of-transit data and black shows all in-transit data +/- 6 hours from the mid-transit time of 2009 Jun 05 06:54 UT. The blue spectrum shows telluric transmission and again the removal of Earth's $O_2$ is imperfect. The upper panel is in-transit/out-of-transit. Stellar wobble is uncorrected here and broadens the out-of-transit spectral lines, leaving excess absorption. There is no planetary signal evident at the HD 80606b's radial velocity, taken to be -59.6 km/s from Colon et al. (2012) and marked by the blue dotted line.

Noise levels per pixel in the Fig. 3 HIRES transit spectrum are 0.8% and upper limits on the depth of potential absorption lines can be set by assuming their linewidth. The transit photometry in Colón *et al.* (2012) measured approximately equal flux in the two blue bands, which contain the K D1 line core and Mg I at 7691.55Å. The $8.09 \pm 2.88 \times 10^{-4}$ flux increase that they reported from these two blue bands to the reddest filter was integrated over OSIRIS' 12Å FWHM bandpass, and so if narrow atomic lines within the HD80606b atmosphere were responsible, they would have to be quite deep. The stellar K D1 linewidth in Fig. 3 is 0.27Å, and at this width, their measurement would correspond to a 3.3% depth if it were due to absorption lines in the blue OSIRIS bands. The HIRES result rules out this possibility; these two datasets would be inconsistent if narrow atomic lines were the responsible absorption mechanism and there is no obvious absorption at the planetary Doppler separation from the stellar K or Mg lines. The noise floor in Fig. 3 could be insufficient to probe very broad features, however. When an ion is neutralized, the radial component of the ion's gyro-motion is preserved in the atom, causing very broad lines (as appear in the following section). An exomoon-magnetosphere interaction could yield very broad and weak absorption features that would be more challenging to distinguish from the noise levels in Fig. 3. Consequently, measurements here do not refute the report by Colón *et al.* (2012) that the HD 80606b system has measureable potassium at high blueshift.

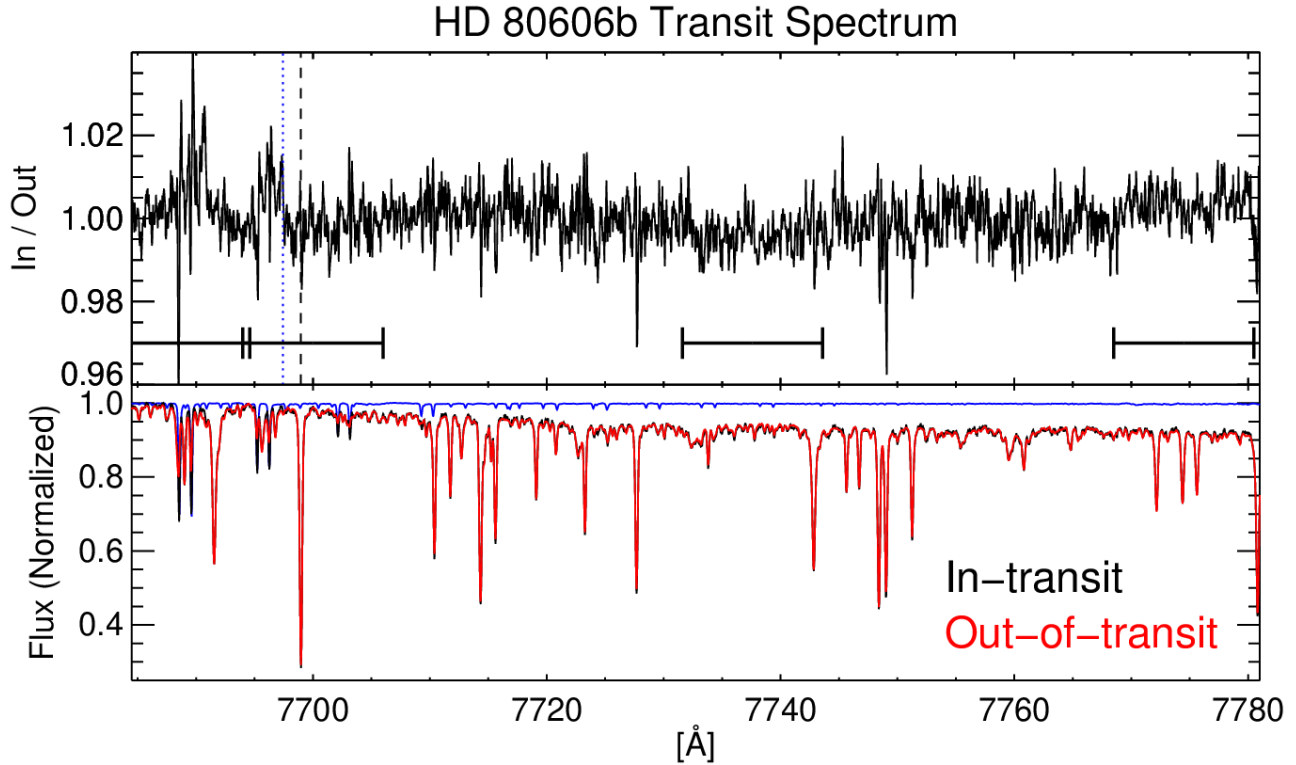

Figure 3. Lower panel: Keck HIRES spectra of HD 80606 timed in (black) and out (red) of the HD 80606b transit. Peak flux has been normalized to unity. The blue spectrum shows the telluric transmission model in-transit. OSIRIS tunable filter bandwidths sampled by Colón *et al.* (2012) are shown as black bars. Upper panel: ratio of in/out of transit. Dashed black and dotted blue lines indicate of the stellar and planetary potassium D wavelengths, respectively (as in Figure 2).

## 5 Fast Na and K in the Io-Jupiter system

The energization mechanism producing fast Na at Io was debated initially and assumed to be an elastic collision or charge exchange, but the directional jet feature almost certainly comes from the reaction $NaCl^+ + e- \rightarrow Na* + Cl*$ (Lellouch *et al.*, 2003; Grava *et al.*, 2014). By analogy with Io, dissociative recombination of alkali bearing molecules from photolysis-born $NaCl^+$ and $KCl^+$ are worth considering as a viable mechanism to energize alkali atoms to the extreme Doppler shifts that exoplanet observers have reported. The ratio of NaCl/KCl is a function of volcanic gas temperature due to their different condensation temperatures (Fegley and Zolotov, 2000). Around close-in planets that are favored by current detection techniques, the tidal forcing on exomoons is expected to melt silicates and hence the Na/K ratio may give some diagnostic of magma temperature in the 1000K to 1500K range. However, the molecular photolysis and recombination efficiency also needs to be accounted for in examining the ratio of atomic Na and K, and exogenic sources like dust or micrometeors may have a non-negligible imprint on composition.

While numerous studies of Io's Na jet have been made, only one measurement of a K jet at Io has been reported (Thomas, 1996). Figure 4 shows measurements of Io's fast alkali streams from Keck HIRES on July 16 2021. A 14″ x 1.7″ aperture was aligned parallel to the centrifugal equator of the Io plasma torus. Pointing was centered at the receding ansa 5.62 Jupiter radii west of Jupiter, where plasma emissions peak just interior of Io's orbit (Schmidt *et al.*, 2018). A spectrum of jovian scattered light, obtained by pivoting off-torus about Jupiter, has been subtracted off to isolate faint alkali D line emissions. Dotted lines show the co-rotational Doppler shift of the jovian magnetosphere at this location. Note that emissions from the jet can appear faster than the magnetospheric plasma flow (e.g. for Na D2) because the ion gyro-motion can add constructively depending on where the recombination occurred within the ion gyro-orbit. A confluence of factors is responsible for the fact that jet distribution is skewed towards lower velocities---not symmetric about the co-rotation velocity. First, the residence time for an atom in the slit is inversely proportional to its transverse velocity. In spatially resolved measurements, faster speeds thus decrease the cloud's density, an explanation Thomas (1996) proposed for the observed potassium velocity distribution. Second, the plasma flow speed could be stagnated local to Io (e.g., Dols *et al.* 2012) and if ion pickup occurs at or below Io's exobase, as Wilson and Schneider (1999) proposed, then ion-neutral collisions could slow these ions, and hence the neutrals they yield at recombination.

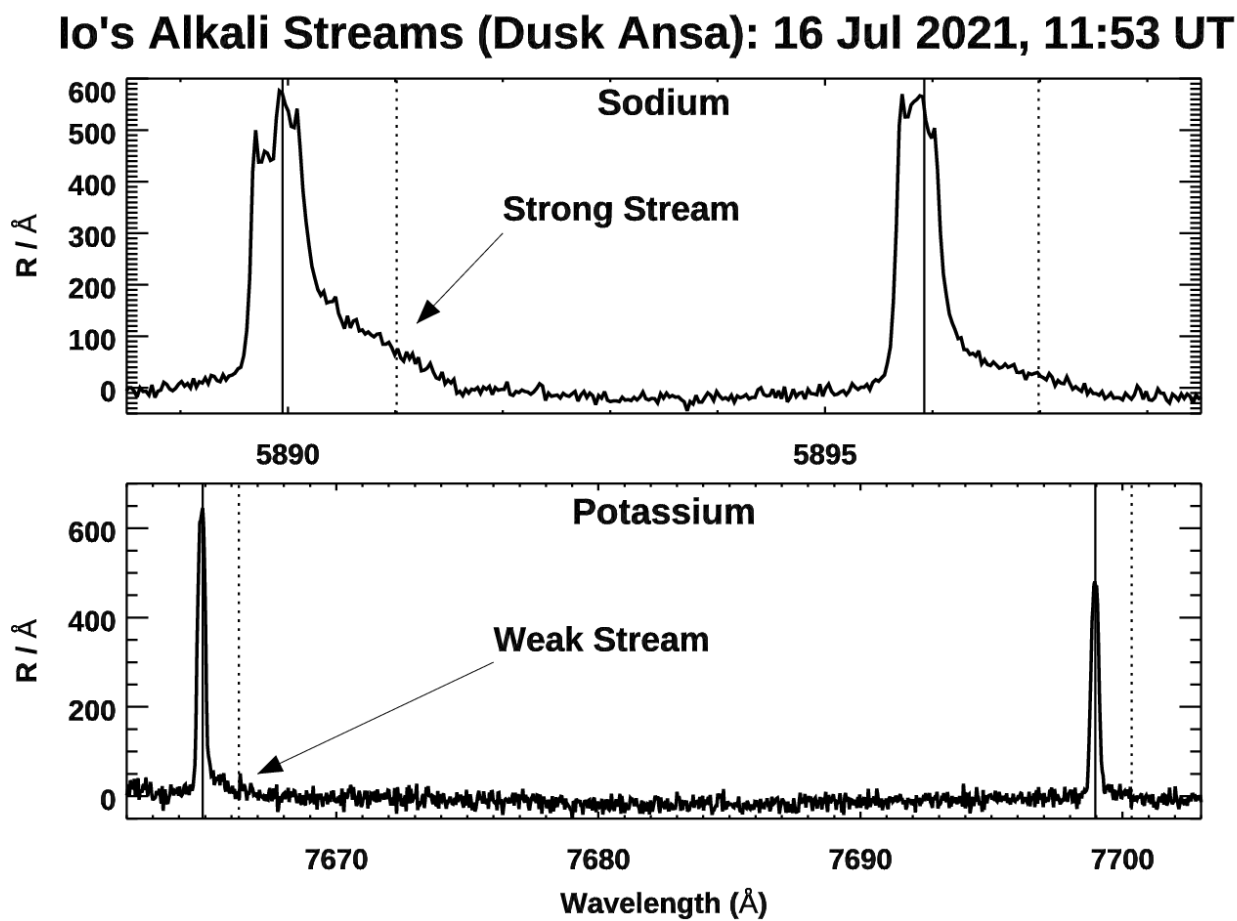

Figure 4. Keck HIRES spectrum of Na and K D emission at the dusk ansa of the Io plasma torus, showing Doppler structure. Spectral flux here has units of Rayleighs/Å. The Keplerian velocity is shown in the solid lines, and the co-rotational velocity of the jovian magnetosphere is shown as the dotted lines. Pointing trails Io, which is at 330° orbital longitude. The red-shifted stream of Na atoms is due to ion neutralization (Na jet). Absence of this feature in the K line profiles suggests the potassium reaction is less efficient.

Accounting for the photon excitation frequency, the ratio of the Na / K column density in Figure 4 is 10.2, consistent with the Brown (2001) estimate in Io's corona. However, the K D lines are far narrower. Compared with Na, K velocities are concentrated at the Keplerian speed (solid lines) and show little of the stream feature produced by ion-neutralization. The K-D line profiles are not symmetric and this result does not refute the Thomas (1996) observation of some potassium at the ion pick-up velocity. However, from the relative structure in the two line profiles, one can conclude that ion neutralization into atomic Na is much more efficient than into atomic K. Consequently, Na D is a more powerful probe for exomoon Doppler signatures relative to K D. At the measured temperatures of Io's escaping corona, the transport time from Io to the region measured here is roughly a day. Photo-ionization lifetimes will be much shorter at transiting exoplanets and consequently, the ratio of the fast stream, relative to the slow cloud would be greater than what is measured here.

## 6     Conclusion

Stellar and planetary absorption lines from alkali metals can be challenging to fully disentangle in transit spectroscopy. Atmospheric signatures in eccentric transiting exoplanets can be more straightforward to interpret since the stellar and planetary lines can appear Doppler shifted. In certain geometries, moon-magnetosphere interactions are capable of producing alkali jets that are strongly shifted from the expected planetary wavelengths. The Doppler shift direction and magnitude depends on the moon's orbital phase. While blue-shifted features are degenerate with expected dynamics from radiation pressure on an exoplanet's escaping exosphere, red-shifted features cannot be attributed to a known physical mechanism, making them a valuable marker. Redshifts have not been reported among the nearly 30 systems that are known to exhibit alkalis in transit spectroscopy. Blue-shifts are commonly reported, and in only one case is the radiation pressure definitively insufficient to account for the magnitude: the fast potassium winds interpreted from transit measurements of HD 80606b (Colón *et al.*, 2012). Follow-up observations and archival analysis could not confirm a broad K signature in transit, but were noise-limited. New measurements of the alkali jets produced by the Io-Jupiter interaction suggest fast K production is inefficient compared to Na. In turn, strongly shifted Na D-line absorptions, particularly in red-shift, may serve as indirect evidence for the presence of both moons and magnetospheres of extrasolar planets, both of which have few proven observational markers at present.

## 7     Author Contributions

Carl Schmidt is the sole contributor to this article, however, the support of several other investigators is acknowledged below.

## 8     Funding

Observation and analysis of the HD 80606b transit at WIYN was supported by NASA under JPL RSA Contract No. 1624307. This work was partially supported by National Science Foundation program AST-2108416.

## 9     Acknowledgments

C.S. thanks Kwangsu Ahn, Dale Gary, and Kevin Reardon for their assistance with Mercury transit observations at Big Bear Solar Observatory, and Luke Moore and Heidi Schweiker for their valuable assistance with the observations at WIYN. Katherine de Kleer, Mike Brown, Maria Camarca and Zachariah Milby also provided assistance with Keck observation of Io's extended neutral clouds. Valuable discussions with Knicole Colón and Aurélien Wyttenbach are also acknowledged. Some of the data presented herein were obtained at the W. M. Keck Observatory, which is operated as a scientific partnership among the California Institute of Technology, the University of California and the National Aeronautics and Space Administration. The Observatory was made possible by the generous financial support of the W. M. Keck Foundation. The authors wish to recognize and acknowledge the very significant cultural role and reverence that the summit of Maunakea has always had within the indigenous Hawaiian community. We are most fortunate to have the opportunity to conduct observations from this mountain.

# 1    Data Availability Statement